\documentclass[11pt]{article}
\usepackage{doublespace}
\usepackage{epsfig}
\usepackage{graphicx}
\usepackage{amssymb}
\usepackage{amsmath}

\begin{document}
\begin{spacing}{1.}
\title{Fractal dimension and self-similarity in \\ {\it Asparagus plumosus} \\
\author{J. R. Castrej\'on Pita,$^1$ A. Sarmiento Gal\'an,$^2$ and R. Castrej\'on Garc{\'\i}a.$^3$ \\
\\
{\small $^1$ Centro de Investigaci\'on en Energ{\'\i}a UNAM, Ap. Postal 34, 62580 Temixco, Morelos,
M\'exico} \\
{\small $^2$ Instituto de Matem\'aticas UNAM, Av. Universidad s/n, 62200 Chamilpa, Morelos, M\'exico} \\
{\small $^3$ Instituto de Investigaciones El\'ectricas, Av. Reforma 113, 62490 Temixco, Morelos,
M\'exico} \\ }}

\maketitle
\end{spacing}

Running head: Fractal dimension in {\it Asparagus plumosus}

\bigskip
\begin{spacing}{1.5}

\begin{abstract}
We measure the fractal dimension of an African plant that is widely
cultivated as ornamental, the {\it Asparagus plumosus}. This plant presents
self-similarity, remarkable in at least two different scalings. In the
following, we present the results obtained by analyzing this plant via the box
counting method for three different scalings. We show in a quantitatively way
that this species is a fractal.
\end{abstract}
\vfill\eject

\section{Introduction}

Nowadays it is frequent to use computational algorithms in order to produce
images of plants and trees that resemble their natural counterparts. These
visualizations, which present several symmetric bifurcations
\cite{Mandelbrot}, encouraged us to analyze the {\it Asparagus plumosus}
\cite{Catalogue}. This plant is a native of Africa, but often cultivated in
the rest of the world as an ornament. The plant can be easily identified: it
is semi-climbing, has a typical height of 2 m, its main branches measure from
25 to 50 cm, and all branches have philiform divisions; its flowers are white
and have six petals each, their fruits are purple spheres, 7 mm in diameter.
Observed in some detail (Figs. 1-3), the `leaves' of this plant, consist of
repeated bifurcations from the main stem, showing a high degree of both,
symmetry and scaling; these branching can also be observed even at the
smallest scale. Two other peculiar characteristics of the `leaves' of this
plant are their flatness and their uniform green color. Although the branches
may be dramatically different in shape (actually, Fig. 2 shows an atypical
branch), we will show that their fractal dimension is the same.

\section{Method}

The method of box counting is widely known \cite{Chaos}. Briefly, the box 
counting technique consists of counting the number of boxes in a grid that 
intersect any part of an image that has been placed over it. In order to 
calculate the fractal dimension of the image, denoted by $D$, using a square 
grid of side size given by $\varepsilon $, one needs to analyze the changes 
in the number of boxes required to cover the image, $N$, as the size of the 
grid is reduced, {\it i. e.},

\begin{equation}
D=\lim_{\varepsilon \rightarrow 0}\frac{\log N(\varepsilon )}{\log 1/\varepsilon }.
\end{equation}

\noindent{We have applied this method to the three bran\-ches shown in Figs. 
1-3 at three different levels: the three different scales at which symmetry is 
observed. In Figs. 3-4, we visually exemplify the application of the box 
counting method. The `leaves' we have designated as medium-size branches 
correspond to the ramifications at the lower right corner of the branches in 
Figs. 1-3, and those called small-size branches were selected from the 
medium-size ones following the same criteria; Fig. 4 exemplifies the selection 
for the main branch in Fig. 3. All the images were obtained by positioning the 
corresponding branch directly on a scanner ($640$ x $460$ resolution, bit map 
images), and since the leaves are objects immersed in a two dimensional space, 
it was not necessary to use any kind of projection. The digital scanning was 
made in black \& white, and in real scale. The side size of the square grid 
was varied from $1$ to $200$ pixels, by steps of 1 pixel. The original size of 
the main branch in Fig. 3 is $428.8 $ x $492.0$ mm, $127.5$ x $220.6$ mm for 
the medium-size branch and $23.0$ x $66.9$ mm for the small one. Since our 
images all have well defined borders, there is no need to analyze the contour 
threshold \cite{Fluids}.}

\section{Results}

The values of $N$, obtained varying the grid size from $1$ to $200$ pixels,
are shown in Figs. 5-7. This pixel range allows for a direct comparison in
real scale of the results for the three levels at which similarity is
observable. A bigger side-size box is not used because the width of the
smallest branches (at the base) is 200 pixels, and therefore, a bigger 
side-size box would mean that a single box would almost cover the whole 
branch. Since the relations are linear over a wide range of $\varepsilon $ 
values, the fractal dimension $D$ is then given by the slope of the 
corresponding line, see Figs. 5-7. Finally, the values obtained for the 
fractal dimension of the three branches and at the three different scales, are 
shown in the Table together with the uncertainty in the slope ($\Delta D$) and 
the correlation of the linear regression ($R$).

\section{Conclusion}

From the previous analysis, where we have shown that the fractal dimension of
the three branches is practically the same, we can conclude that the shape of
a branch of {\it Asparagus plumosus} is independent for the determination of
its fractal dimension. The very small uncertainties in these values
($\Delta D/D < 3$ x $10^{-3}$) can be easily interpreted in terms of the high
linear correlations shown in the Table. Accordingly, we can confirm the
fractal dimension in this species, a new type of natural fractal being added
to the extensive already well know gallery (for a recent, man-made example,
see \cite{Tokyo}). Additionally, since the value of the fractal dimension
obtained from the analysis of the two bigger scales is indeed very similar,
we can conclude that there is the same level of complexity at these two
scales: the plant is self-similar. Unfortunately, we do not seem to find the
same self-similarity at the smallest scale.

\end{spacing}
\begin{spacing}{.5}
\footnote{Martin Nezadal and Oldrich Zmeskal (Institute of Physical and
Applied Chemistry at Brno, Czech Republic) are greatefully aknowledged for
their HarFA program. This work has been partially supported by DGAPA-UNAM
(IN101100), and UC-MEXUS.}
\end{spacing}

\vfill\eject

\bigskip
\centerline{Table}
\begin{tabular}[t]{c}
\hline\hline \qquad \qquad Fractal dimension ($D$) \qquad Uncertainty ($\Delta D$) \qquad Linear regression ($R$) \\ \hline
\begin{tabular}{l}
Branch \ \ \\
Fig. 1 \\
Fig. 2 \\
Fig. 3
\end{tabular}
\begin{tabular}{lllllllll}
Main & Med & Small \ \ & Main & Med & Small & \ \ Main & Med & Small\\
1.742 & 1.712 & 1.825 & 0.003 & 0.003 & 0.005 & \ \ 0.999 & 0.999 & 0.999\\
1.787 & 1.765 & 1.869 & 0.002 & 0.002 & 0.005 & \ \ 0.999 & 0.999 & 0.999\\
1.760 & 1.722 & 1.819 & 0.002 & 0.003 & 0.006 & \ \ 0.999 & 0.999 & 0.998\\
\end{tabular}
\\ \hline
\end{tabular}
\vfill\eject

\begin{figure}[tbp]
\begin{center}
\epsfig{width=10cm,file=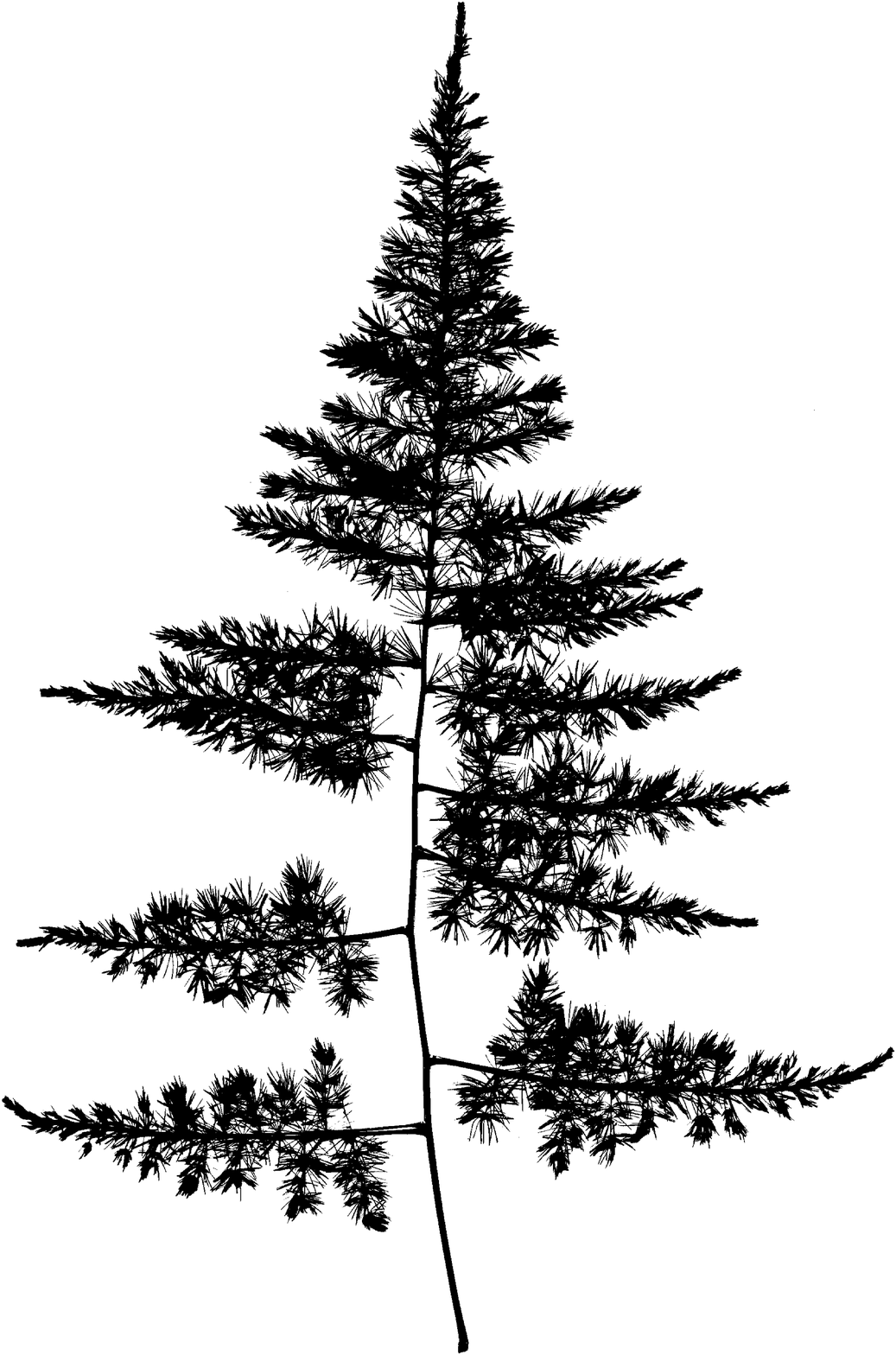}
\par
Fig. 1  A typical example of a main branch of \textit{Asparagus plumosus}.
\end{center}
\end{figure}
\vfill\eject

\begin{figure}[tbp]
\begin{center}
\epsfig{width=10cm,file=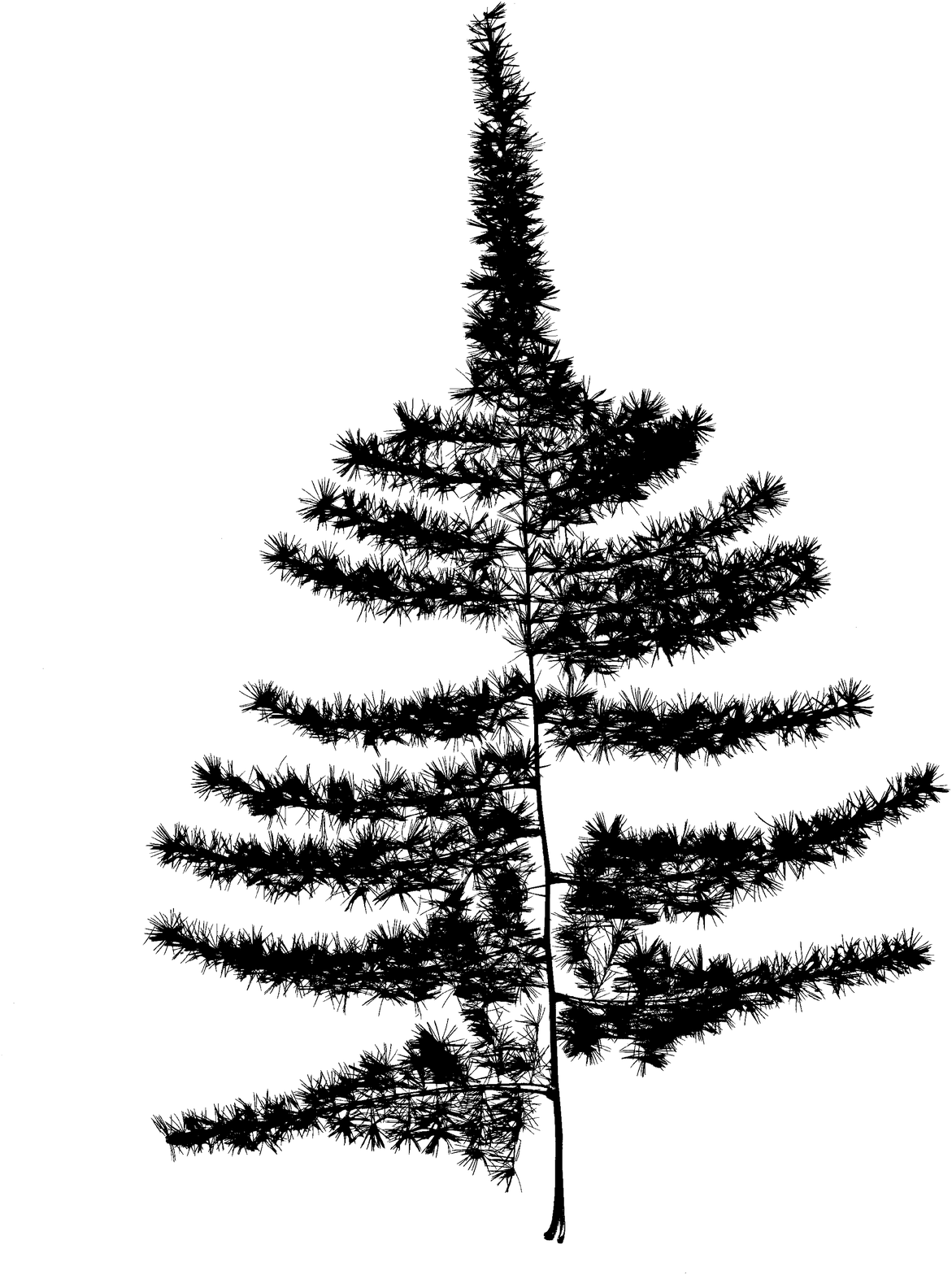}
\par
Fig. 2  An atypical main branch of \textit{Asparagus plumosus}, note the differences in shape with
respect to the usual branches in Figs. 1 and 3
\end{center}
\end{figure}
\vfill\eject

\begin{figure}[tbp]
\begin{center}
\epsfig{width=10cm,file=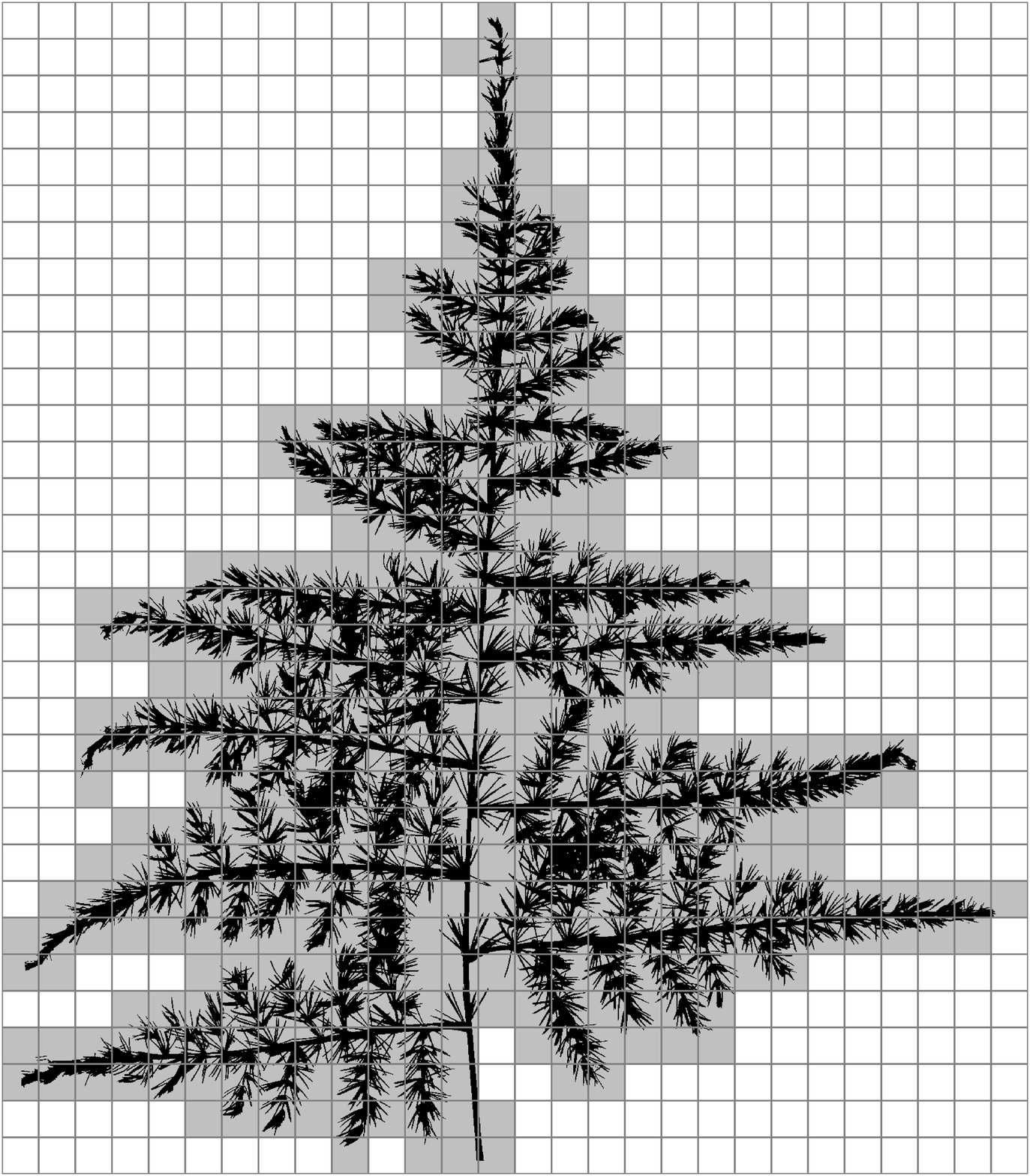}
\par
Fig. 3  Third example of a main branch; a grid of boxes with a side-length of $60$ pixels is also
shown (those boxes that have an intersection with the image are shaded in gray).
\end{center}
\end{figure}
\vfill\eject

\begin{figure}[tbp]
\begin{center}
\epsfig{width=10cm,file=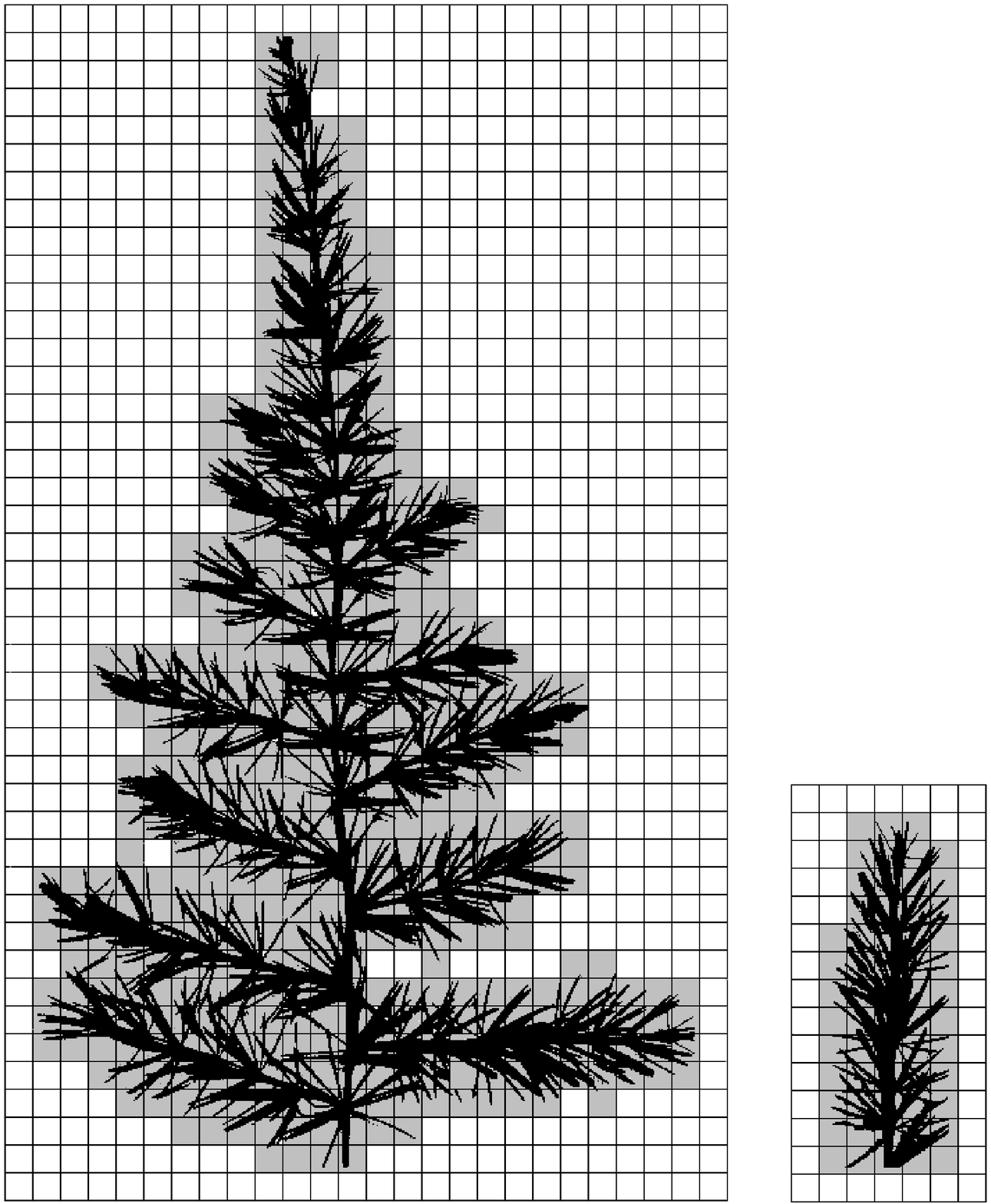}
\par
Fig. 4  Medium- and small-size branches lying on a square grid with side-length of $20$ pixels.
\end{center}
\end{figure}
\vfill\eject

\begin{figure}[tbp]
\begin{center}
\epsfig{width=12cm,file=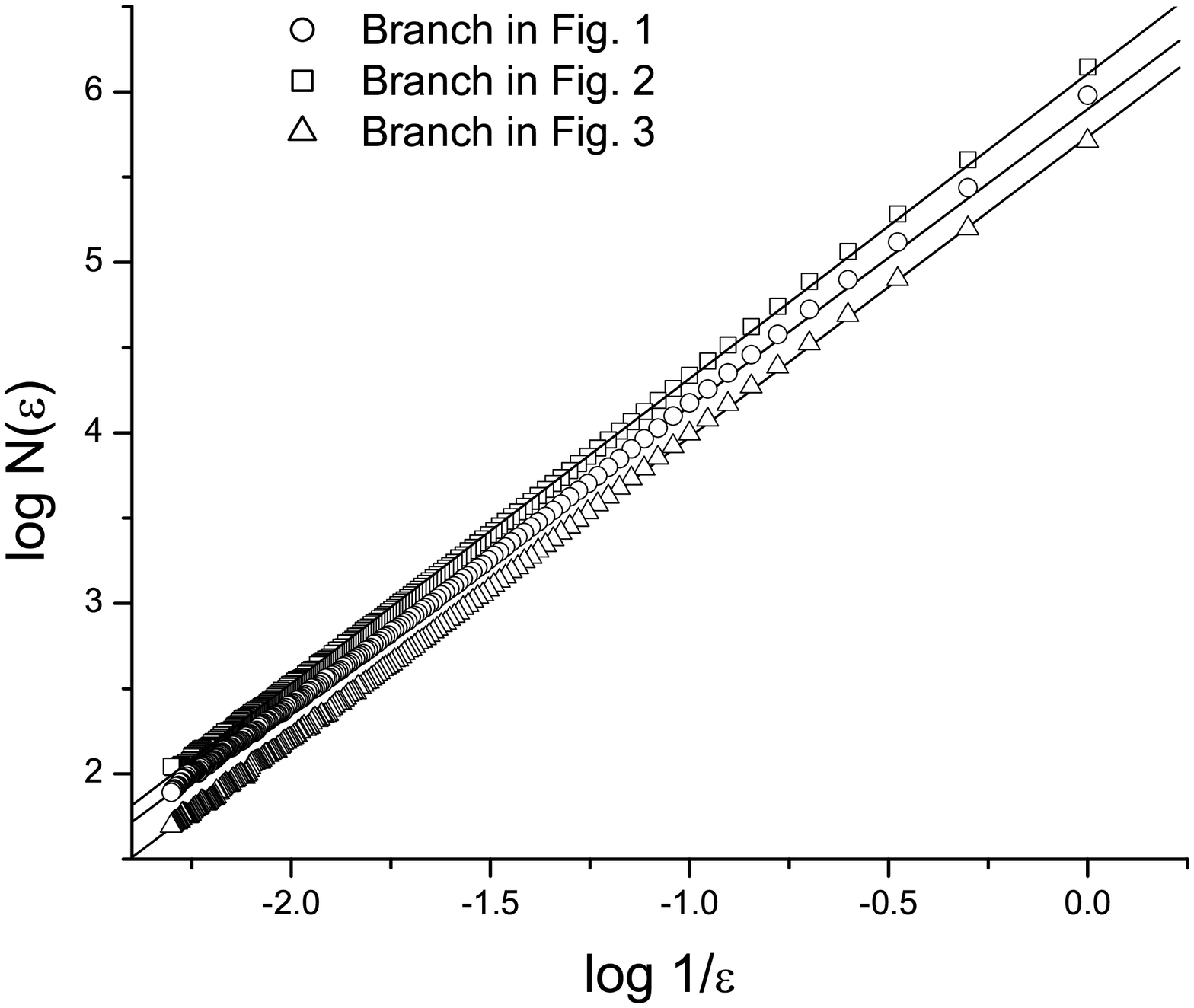}
\par
Fig. 5  Symbols represent the results of applying the box counting method to the main branches in
Figs. 1-3. Straight lines show linear regressions performed for each data set.
\end{center}
\end{figure}
\vfill\eject

\begin{figure}[tbp]
\begin{center}
\epsfig{width=12cm,file=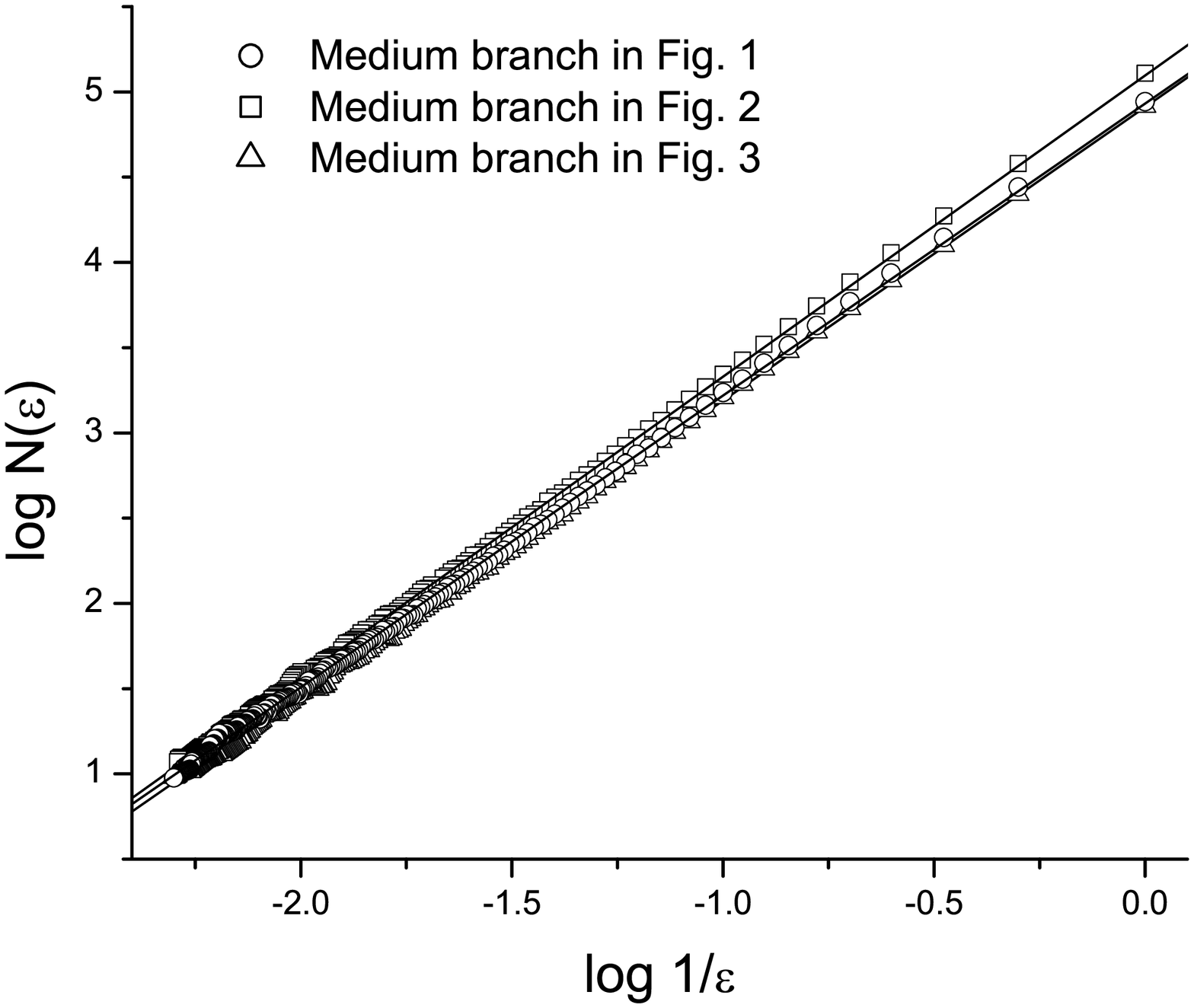}
\par
Fig. 6  Analogue of Fig. 5 for medium-size branches.
\end{center}
\end{figure}
\vfill\eject

\begin{figure}[tbp]
\begin{center}
\epsfig{width=12cm,file=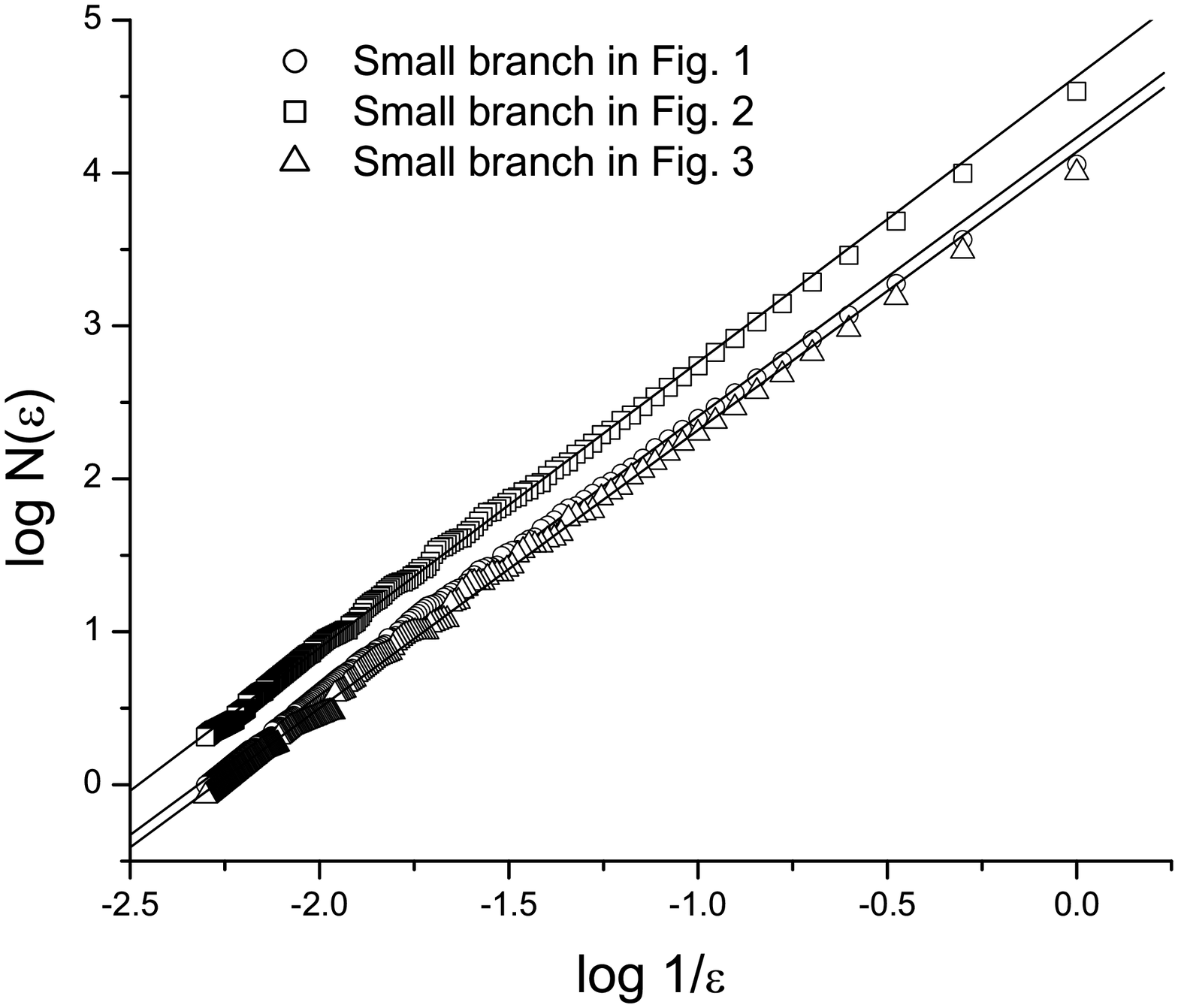}
\par
Fig. 7  Analogue of Fig. 5 for small-size branches.
\end{center}
\end{figure}
\vfill\eject


\begin{thebibliography}{9}

\bibitem{Mandelbrot}  B. B. Mandelbrot, {\it The Fractal Geometry of Nature} (Freeman, San Francisco,
1989).

\bibitem{Catalogue}  M. Mart{\'\i}nez, {\it Cat\'alogo de Nombres Vulgares y Cient{\'\i}ficos de
Plantas Mexicanas} (Fondo de Cultura Econ\'omica, M\'exico, 1979).

\bibitem{Chaos}  K. T. Alligood, T. D. Sauer and J. A. Yorke, {\it Chaos: An In\-tro\-duc\-tion to
Dy\-na\-mi\-cal Sys\-tems}, (Springer Press: New York, 1996), pp 172-180.

\bibitem{Fluids}  R. R. Prasad and K. R. Sreenivasan, {\it Phys. Fluids A}, {\bf 2}, 5 (1990).

\bibitem{Tokyo}  V. Rodin and E. Rodina, {\it Fractals}, {\bf 8}, 4 (2000).

\end{thebibliography}
\end{document}